\documentstyle[a4,11pt]{article}
\begin{document}

\begin{titlepage}
\vskip 2cm
\begin{flushright}
Preprint CNLP-1995-06
\end{flushright}
\vskip 2cm
\begin{center}
{\large {\bf ON SOME SIGMA MODELS WITH POTENTIALS AND THE KLEIN - GORDON TYPE
EQUATIONS}}\footnote{Preprint
CNLP-1995-06.Alma-Ata.1995 \\
cnlpmyra@satsun.sci.kz}
\vskip 2cm

{\bf Ratbay Myrzakulov  }

\end{center}
\vskip 1cm

$^{a}$ Centre for Nonlinear Problems, PO Box 30, 480035, Almaty-35, Kazakstan

\begin{abstract}
The gauge equivalent counterparts of the some (1+1)-, or (2+0)-
dimensional $\sigma$-models are found.
Also we have proved the equivalence between the some spin-phonon
equations and the Yajima-Oikawa-Ma equations.
\end{abstract}


\end{titlepage}

\setcounter{page}{1}
\newpage

\tableofcontents
\section{Introduction}

Integrable $\sigma$ - models
play important role in the modern theoretical and mathematical physics.
They arise in gravity theory, extended supergravity, in theory of
Anderson localization, the Kaluza-Klein theory, in theory of strings,
and superstrings. The siplest nonrelativistic version of the $\sigma$-model
is a continuous classical spin Heisenberg model or the (2+1)-dimensional
isotropic Landau-Lifshitz equation(LLE)
$$
{\bf S}_{t} = {\bf S}\wedge \triangle {\bf S}
\eqno (1)
$$
and its extensions to higher spins and to multidimensions[1]. Then in
stationary  limit the LLE (1) coincide with the $\sigma$-model equation
$$
 \triangle {\bf S} + (\nabla {\bf S})^{2} {\bf S} = 0. \eqno (2)
$$
Here
$$
{\bf S}^{2} = S^{2}_{3} + r^{2} (S^{2}_{1} + S^{2}_{2}) = E = \pm 1, \eqno(3)
$$
$$
 \triangle = \frac{\partial^{2}}{\partial x^{2}} +
\alpha^{2} \frac{\partial^{2}}{\partial y^{2}},
\quad
\nabla = {\bf i}\frac{\partial}{\partial x} +
{\bf j}\frac{\partial}{\partial y}, \quad {\bf i}^{2} = 1, \quad {\bf j}^{2}
 = \alpha^{2},  \quad \alpha^{2} = \pm 1, \quad
r^{2} = \pm 1, \quad E = \pm 1. \eqno (4)
$$

In this work we consider the some generalizations of the $\sigma$-model (2),
namely, the $\sigma$-models with potentials.

Let us now we consider the Myrzakulov IX (M-IX) equation [1]
$$ iS_t+\frac{1}{2}[S,M_1S]+A_2S_x+A_1S_y = 0 \eqno(5a)$$
$$ M_2u=\frac{\alpha^{2}}{2i}tr(S[S_x,S_y]) \eqno(5b)$$
where $ \alpha,b,a  $=  consts and
$$
S= \pmatrix{
S_3 & rS^- \cr
rS^+ & -S_3
},\quad S^{\pm}=S_{1}\pm iS_{2} \quad  S^2 = EI,\quad r^{2}=\pm 1,
$$
$$
M_1= \alpha ^2\frac{\partial ^2}{\partial y^2}+4\alpha (b-a)\frac{\partial^2}
   {\partial x \partial y}+4(a^2-2ab-b)\frac{\partial^2}{\partial x^2},
$$
$$
M_2=\alpha^2\frac{\partial^2}{\partial y^2} -2\alpha(2a+1)\frac{\partial^2}
   {\partial x \partial y}+4a(a+1)\frac{\partial^2}{\partial x^2},
$$
$$
A_1=i\{\alpha (2b+1)u_y - 2(2ab+a+b)u_{x}\},
$$
$$
A_2=i\{4\alpha^{-1}(2a^2b+a^2+2ab+b)u_x - 2(2ab+a+b)u_{y}\}.
$$

This set of equations arises from the compatibility condition of the
following linear equations[1]
$$ \alpha \Phi_y =\frac{1}{2}[S+(2a+1)I]\Phi_x \eqno(6a) $$
$$ \Phi_t=2i[S+(2b+1)I]\Phi_{xx}+W\Phi_x \eqno(6b) $$
with
$$ W = 2i\{(2b+1)(F^{+} + F^{-} S) +(F^{+}S + F^{-}) +
(2b-a+\frac{1}{2})SS_x+\frac{1}{2}S_{x} + \frac{\alpha}{2} SS_y \}, \quad
$$
$$
F^{\pm} = A \pm D , \quad A=i[u_{y} - \frac{2a}{\alpha} u_{x}],
\quad D=i[\frac{2(a+1)}{\alpha} u_{x} - u_{y}].
$$

It is well known that equation (5) is gauge and Lakshmanan equivalent
to the following Zakharov equation(ZE)[2]
$$
iq_{t}+M_{1}q+vq=0, \eqno(7a)
$$
$$
ip_{t}-M_{1}p-vp=0, \eqno(7b)
$$
$$
M_{2}v = -2M_{1}(pq), \eqno(7c)
$$

The Lax representation of this equation has the form
$$
\alpha \Psi_y =B_{1}\Psi_x + B_{0}\Psi, \eqno(8a)
$$
$$
\Psi_t=iC_{2}\Psi_{xx}+C_{1}\Psi_x+C_{0}\Psi, \eqno(8b)
$$
with
$$
B_{1}= \pmatrix{
a+1 & 0 \cr
0   & a
},\quad
B_{0}= \pmatrix{
0   &  q \cr
p   &  0
}
$$
$$
C_{2}= \pmatrix{
b+1 & 0 \cr
0   & b
},\quad
C_{1}= \pmatrix{
0   &  iq \cr
ip  &  0
},\quad
C_{0}= \pmatrix{
c_{11}  &  c_{12} \cr
c_{21}  &  c_{22}
}
$$
$$
c_{12}=i(2b-a+1)q_{x}+i\alpha q_{y},\quad
c_{21}=i(a-2b)p_{x}-i\alpha p_{y}.
$$
Here $c_{jj}$ is the solution of the  following  equations
$$
(a+1) c_{11x}- \alpha c_{11y} = i[(2b-a+1)(pq)_{x} + \alpha (pq)_{y}],
\quad ac_{22x}-\alpha c_{22y} = i[(a-2b)(pq)_{x} - \alpha (pq)_{y}]. \eqno (9)
$$

The goal of this work is to explore the some $\sigma$-models with potentials
derivable from the M-IX equation(5).

\section{Gauge equivalence between the some $\sigma$-models with potentials
and  the Klein-Gordon-type equations}

It is interesting  to note that the equation(5) admits  some  reductions
in 1+1 or 2+0 dimensions. Let us now consider these reductions.

\subsection{The Myrzakulov XXXII (M-XXXII) equation}

Suppose that now $\nu = t$ is the some "hidden" parameter, $S = S(x,y), \quad
u = u(x,y), \quad  q = q(x,y), \quad p = p(x,y), \quad v = v(x,y) $ and
at the same time $\Phi = \Phi(x,y,\nu), \quad
\Psi = \Psi(x,y,\nu)$.
Then equation (5) reduces to the following $\sigma$-model with potential
$$
M_1S + \{ k_{1} S^{2}_{x} + k_{2} S_{x}S_{y} + k_{3} S^{2}_{y} \} S
+ A_2 SS_x+A_1 SS_y = 0 \eqno(10a)$$
$$ M_2u=\frac{\alpha^{2}}{2i}tr(S[S_x,S_y]) \eqno(10b)$$
where $M_{1}$ we write in the form
$$
M_1= k_{3}\frac{\partial ^2}{\partial y^2}+k_{2}\frac{\partial^2}
   {\partial x \partial y}+k_{1}\frac{\partial^2}{\partial x^2},
$$
which is the compatibility condition $ \Phi_{y\nu} = \Phi_{\nu y}$
of the set (6) and called the M-XXXII equation[1]. The G-equivalent
and L-equivalent counterpartt of this equation is given by
$$
M_{1}q+vq=0, \quad M_{1}p + vp=0, \quad M_{2}v = -2M_{1}(pq). \eqno(11)
$$
This is the some modified complex Klein-Gordon equation (mKGE).

\subsection{The Myrzakulov XV (M-XV) equation}

Let us consider the case: $a = b$. Then we obtain the M-XV equation [1]
$$
\alpha^{2} S_{yy} - a(a+1) S_{xx} +
\{\alpha^{2} S^{2}_{y} - a(a+1) S^{2}_{x}\} S +
+  A_{2}^{\prime \prime } SS_x + A_{1}^{\prime \prime } SS_y  \eqno(12a)
$$
$$
M_{2} u = \frac{\alpha^{2}}{2i} tr( S [S_{x}, S_{y}]) \eqno(12b)
$$
where $A_{j}^{\prime \prime} = A_{j}$ as $ a = b$. The corressponding mKGE
has the form
$$
\alpha^{2}q_{yy} - a(a+1)q_{xx} + v q = 0, \quad
M_{2} v = -2[\alpha^{2} (\mid  q \mid^{2})_{yy} -
a(a+1) (\mid q\mid^{2})_{xx}] \eqno(13)
$$
The Lax representations of these equations we obtain from (6) and (8)
respectively as $a = b$.

\subsection{The Myrzakulov XIV (M-XIV) equation}

Now we  consider the reduction: $a=-\frac{1}{2}$. Then the equation (10)
reduces to the M-XIV equation[1]
$$
S_{xx} + 2\alpha(2b+1)S_{xy}+\alpha^{2}S_{yy}) +
\{S_{xx} + 2\alpha(2b+1)S_{xy}+\alpha^{2}S_{yy}\}S +
A^{\prime}_{2}SS_x+A^{\prime}_{1}SS_y = 0 \eqno(14a)
$$
$$
\alpha^{2}u_{yy} - u_{xx}=
\frac{\alpha^{2}}{2i}tr(S[S_x,S_y]) \eqno(14b)
$$
where $A^{\prime}_{j} = A_{j}$ as $a=-\frac{1}{2}$. The corresponding gauge
equivalent equation is obtained from (11) and looks like
$$
q_{xx} + 2\alpha(2b+1)q_{xy}+\alpha^{2}q_{yy}+vq = 0 \eqno(15a)
$$
$$
\alpha^{2}v_{yy} - v_{xx}=-2\{\alpha^{2}(pq)_{yy} +
2\alpha (2b+1)(pq)_{xy} +(pq)_{xx}\} \eqno(15b)
$$
From (6) and (8) we obtain the Lax representations of (14) and (15)
respectively as $ a = -\frac{1}{2}$.

\subsection{The Myrzakulov XIII (M-XIII) equation}

Now let us consider the case: $ a=b=-\frac{1}{2} $. In this case the
equations (10) reduce to the $\sigma$-model
$$
S_{xx}+\alpha^{2}S_{yy} +
\{S^{2}_{x}+\alpha^{2}S^{2}_{y}\}S +
iu_{y}SS_x+iu_{x}SS_y = 0 \eqno(16a)
$$
$$
\alpha^{2}u_{yy} - u_{xx} = \frac{\alpha^{2}}{2i}tr(S[S_y,S_x]) \eqno(16b)
$$
which is the M-XIII equation[1].
The  equivalent counterpart of the equation(16) is the following  equation
$$
q_{xx}+\alpha^{2}q_{yy}+ vq = 0 \eqno(17a)
$$
$$
\alpha^{2}v_{yy} - v_{xx}=-2\{\alpha^{2}(pq)_{yy}
 +(pq)_{xx}\} \eqno(17b)
$$
that follows from the mKGE(11).

\subsection{The Myrzakulov XII (M-XII) equation}

Now let us consider the case: $ a=b=-1$. In this case the
equations(10) reduce to the $\sigma$-model - the M-XII equation[1]
$$
S_{YY} +  S^{2}_{Y} S + iw SS_{Y} = 0 \eqno(18a)
$$
$$
w_{Y} + w_{X} +\frac{1}{4i}tr(S[S_X,S_Y]) \eqno(18b)
$$
where $X=2x, \quad Y = \alpha y, \quad w = -\frac{u_{Y}}{\alpha}$.
The  equivalent counterpart of the equation(18) is the mKGE
$$
q_{YY}+ vq = 0 \eqno(19a)
$$
$$
v_{X} + v_{Y} + 2(pq)_{Y} = 0. \eqno(19b)
$$
that follows from the (11).

\subsection{The Myrzakulov XXXI (M-XXXI) equation}

This $\sigma$ -model equation is read as[1]
$$
bS_{\eta \eta} - (b+1) S_{\xi \xi}] +
\{bS_{\eta \eta} - (b+1) S_{\xi\xi}\} S +
i(b+1)w_{\eta} SS_{\eta} + ibw_{\xi}SS_{\xi} =0 \eqno(20a)
$$
$$
w_{\xi \eta} = - \frac{1}{4i}tr(S[S_{\eta},S_{\xi}]) \eqno(20b)
$$
which is the M-XXXI equation, where $ w = -\alpha^{-1} u$.
The   equivalent mKGE looks like
$$
(1 + b)q_{\xi \xi } - b q_{\eta \eta } + vq = 0 \eqno(21a)
$$
$$
v_{\xi \eta } = -2\{(1+ b) (pq)_{\xi \xi} - b(pq)_{\eta \eta}\} \eqno(21b)
$$

So we have found the G-equivalent counterparts of
the $\sigma$-models with potentials.

\section{The other (1+1)-dimensional reductions: gauge equivalence
between the  M$^{53}_{00}$-equation and the Yajima-Oikawa -Ma equation}

Let us now we consider the reduction of the M-IX equation (5) as $ a=b=-1$.
We have
$$
iS_t+\frac{1}{2}[S,S_{YY}]+iwS_Y = 0 \eqno(22a)
$$
$$
w_{X} + w_{Y} + \frac{1}{4i}tr(S[S_x,S_y]) = 0 \eqno(22b)
$$
This equation is the Myrzakulov VIII (M-VIII) equation[1].  The G-equivalent
and L-equivalent counterpart of equation (22) is given by
$$
iq_{t}+ q_{YY} + vq = 0, \eqno(23a)
$$
$$
ip_{t}-p_{YY} - vp = 0, \eqno(23b)
$$
$$
v_{X} + v_{Y} + 2(pq)_{Y} =0.  \eqno(23c)
$$

Now let us take the case when $ X = t$. Then the M-VIII equation (22) pass
to the following M$^{53}_{00}$ - equation [1]
$$
iS_t+\frac{1}{2}[S,S_{YY}]+iwS_Y = 0 \eqno(24a)
$$
$$
w_{t} + w_{Y} + \frac{1}{4}\{tr(S^{2}_{Y})\}_{Y} = 0 \eqno(24b)
$$

The  M$^{53}_{00}$ - equation (24)
was proposed in [1] to describe a nonlinear dynamics of the
compressible magnets (see, Appendix). It is integrable and has  the different soliton
solutions[1].

In our case  equation (23)  becomes
$$
iq_{t}+ q_{YY} + vq = 0, \eqno(25a)
$$
$$
ip_{t}-p_{YY} - vp = 0, \eqno(25b)
$$
$$
v_{t} + v_{Y} + 2(pq)_{Y} =0.  \eqno(25c)
$$
that is the Yajima-Oikawa equation(YOE)[3]. So we have proved that the
M$^{53}_{00}$ - equation (24) and the YOE (25) is gauge equivalent
to each other.  The Lax representations of (24) and (25) we can get from
(6) and (8)
respectively as $ a = b = - 1$( see, for example, the ref.[1]). Note that
our Lax representation for the YOE (25) is different than that which was
presented in [3].

Also we would like note that the M-VIII equation (22) we usually write in
the following form
$$
iS_t=\frac{1}{2}[S_{\xi\xi},S]+iwS_{\xi}   \eqno (26a)
$$
$$
w_{\eta}= \frac{1}{4i}tr(S[S_{\eta},S_{\xi}])  \eqno (26b)
$$

The gauge equivalent counterpart
of this  equation is the following ZE[2]
$$
iq_{t}+q_{\xi \xi}+vq=0, \eqno(27a)
$$
$$
v_{\eta} = -2r^{2}(\bar q q)_{\xi}. \eqno(27b)
$$

As $ \eta = t$  equation (26) take the other form of the M$^{53}_{00}$-
equation
$$
iS_t=\frac{1}{2}[S_{\xi\xi},S]+iwS_{\xi}   \eqno (28a)
$$
$$
w_{t}= \frac{1}{4i}\{tr(S^{2}_{\xi})\}_{\xi}  \eqno (28b)
$$

Similarly, equation (27) becomes
$$
iq_{t}+q_{\xi \xi}+vq=0, \eqno(29a)
$$
$$
v_{t} = -2r^{2}(\bar q q)_{\xi}, \eqno(29b)
$$

which is called the Ma equation and  was considered in [4].

\section{Conclusion}

We have established the gauge equivalence between the (1+1)-, or (2+0)-
dimensional $\sigma$-models and the Klein-Gordon type equations.
Also we have proved the equivalence between the some spin-phonon
equations and the Yajima-Oikawa-Ma equations.

\section{Appendix: On some soliton equations of compressible magnets}

Solitons in magnetically ordered crystals have been widely investigated
from both theoretical and experimental points of view. In particular, the
existence of coupled magnetoelastic solitons in the Heisenberg compressible
spin chain has been extensively demonstrated. In [1]
were presented a new classes of integrable and nonintegrable soliton
equations of spin systems. Below we present the some of these  nonlinear
models of magnets - the some of the Myrzakulov equations(ME), which
describe the nonlinear dynamics of compressible magnets.

\subsection{The 0-class of spin-phonon systems}

The Myrzakulov equations with the potentials have the form:\\
the $ M^{10}_{00}$ - equation:
$$ 2iS_t=[S,S_{xx}]+(u+h)[S,\sigma_3]  $$
the $ M^{20}_{00}$ - equation:
$$ 2iS_t=[S,S_{xx}]+(uS_3+h)[S,\sigma_3]  $$
the $ M^{30}_{00}$ - equation:
$$ 2iS_t=\{(\mu \vec S^2_x-u+m)[S,S_x]\}_x+h[S,\sigma_3]  $$
the $ M^{40}_{00}$ - equation:
$$ 2iS_t=n[S,S_{xxxx}]+2\{(\mu \vec S^2_x-u+m)[S,S_x]\}_x+
h[S,\sigma_3] $$
the $ M^{50}_{00}$ - equation:
$$ 2iS_t=[S,S_{xx}]+2uS_x  $$
where $v_{0}, \mu, \lambda, n, m, a, b, \alpha, \beta, \rho, h$ are constants,
$u$ is a scalar function(potential),
subscripts denote partial differentiations, $[,]$ (\{,\}) is
commutator (anticommutator),
$$S= \pmatrix{
S_3 & rS^- \cr
rS^+ & -S_3
}, \,\,\,\,\, S^{\pm}=S_{1}\pm i S_{2},\,\,\,\, r^{2}=\pm 1\,\,\,\,\, S^2=I.    $$

\subsection{The 1-class of spin-phonon systems}

The $ M^{11}_{00}$ - equation:
$$
2iS_t=[S,S_{xx}]+(u+h)[S,\sigma_3]
$$
$$
\rho u_{tt}=\nu^2_0 u_{xx}+\lambda(S_3)_{xx}
$$
the $ M^{12}_{00}$ - equation:
$$
2iS_t=[S,S_{xx}]+(u+h)[S,\sigma_3]
$$
$$
\rho u_{tt}=\nu^2_0 u_{xx}+\alpha(u^2)_{xx}+\beta u_{xxxx}+
    \lambda(S_3)_{xx}
$$
the $ M^{13}_{00}$ - equation:
$$
2iS_t=[S,S_{xx}]+(u+h)[S,\sigma_3]
$$
$$
u_t+u_x+\lambda(S_3)_x=0
$$
the $ M^{14}_{00}$ - equation:
$$
2iS_t=[S,S_{xx}]+(u+h)[S,\sigma_3]
$$
$$
u_t+u_x+\alpha(u^2)_x+\beta u_{xxx}+\lambda(S_3)_x=0
$$

\subsection{The 2-class of spin-phonon systems}

The $ M^{21}_{00}$ - equation:
$$
2iS_t=[S,S_{xx}]+(uS_3+h)[S,\sigma_3]
$$
$$
\rho u_{tt}=\nu^2_0 u_{xx}+\lambda(S^2_3)_{xx}
$$
the $ M^{22}_{00}$ - equation:
$$
2iS_t=[S,S_{xx}]+(uS_3+h)[S,\sigma_3]
$$
$$
\rho u_{tt}=\nu^2_0 u_{xx}+\alpha(u^2)_{xx}+\beta u_{xxxx}+
\lambda (S^2_3)_{xx}
$$
the $ M^{23}_{00}$ - equation:
$$
2iS_t=[S,S_{xx}]+(uS_3+h)[S,\sigma_3]
$$
$$
u_t+u_x+\lambda(S^2_3)_x=0
$$
the $ M^{24}_{00}$ - equation:
$$
2iS_t=[S,S_{xx}]+(uS_3+h)[S,\sigma_3]
$$
$$
u_t+u_x+\alpha(u^2)_x+\beta u_{xxx}+\lambda(S^2_3)_x=0
$$

\subsection{The 3-class of spin-phonon systems}

The $ M^{31}_{00}$ - equation:
$$
2iS_t=\{(\mu \vec S^2_x - u +m)[S,S_x]\}_x
$$
$$
\rho u _{tt}=\nu^2_0 u_{xx}+\lambda(\vec S^2_x)_{xx}
$$
the $M^{32}_{00}$ - equation:
$$
2iS_t=\{(\mu \vec S^2_x - u +m)[S,S_x]\}_x
$$
$$
\rho u _{tt}=\nu^2_0 u_{xx}+\alpha (u^2)_{xx}+\beta u_{xxxx}+ \lambda
(\vec S^2_x)_{xx}
$$
the  $M^{33}_{00}$ - equation:
$$
2iS_t=\{(\mu \vec S^2_x - u +m)[S,S_x]\}_x
$$
$$
u_t+u_x +\lambda (\vec S^2_x)_x = 0
$$
the  $M^{34}_{00}$ - equation:
$$
2iS_t=\{(\mu \vec S^2_x - u +m)[S,S_x]\}_x
$$
$$
u_t+u_x +\alpha(u^2)_x+\beta u_{xxx}+\lambda (\vec S^2_x)_{x} = 0
$$

\subsection{The 4-class of spin-phonon systems}

The  $M^{41}_{00}$ - equation:
$$
2iS_t=[S,S_{xxxx}]+2\{((1+\mu)\vec S^2_x-u+m)[S,S_x]\}_{x}
$$
$$
\rho u_{tt}=\nu^2_0 u_{xx}+\lambda (\vec S^2_x)_{xx}
$$
the  $M^{42}_{00}$ - equation:
$$
2iS_t=[S,S_{xxxx}]+2\{((1+\mu)\vec S^2_x-u+m)[S,S_x]\}_{x}
$$
$$
\rho u_{tt}=\nu^2_0 u_{xx}+\alpha(u^2)_{xx}+\beta u_{xxxx}+\lambda
(\vec S^2_x)_{xx}
$$
the  $M^{43}_{00}$ - equation:
$$
2iS_t=[S,S_{xxxx}]+2\{((1+\mu)\vec S^2_x-u+m)[S,S_x]\}_{x}
$$
$$
u_t + u_x + \lambda (\vec S^2_x)_x = 0
$$
the  $M^{44}_{00}$ - equation:
$$
2iS_t=[S,S_{xxxx}]+2\{((1+\mu)\vec S^2_x-u+m)[S,S_x]\}_{x}
$$
$$
u_t + u_x + \alpha(u^2)_x + \beta u_{xxx}+\lambda (\vec S^2_x)_x = 0
$$

\subsection{The 5-class of spin-phonon systems}

The  $M^{51}_{00}$ - equation:
$$ 2iS_t=[S,S_{xx}]+2uS_x  $$
$$
\rho u_{tt}=\nu^2_0 u_{xx}+\lambda (f)_{xx}
$$
the  $M^{52}_{00}$ - equation:
$$ 2iS_t=[S,S_{xx}]+2uS_x  $$
$$
\rho u_{tt}=\nu^2_0 u_{xx}+\alpha(u^2)_{xx}+\beta u_{xxxx}+\lambda
(f)_{xx} \
$$
the  $M^{53}_{00}$ - equation:
$$ 2iS_t=[S,S_{xx}]+2uS_x  $$
$$
u_t + u_x + \lambda (f)_x = 0
$$
the  $M^{54}_{00}$ - equation:
$$ 2iS_t=[S,S_{xx}]+2uS_x  $$
$$
u_t + u_x + \alpha(u^2)_x + \beta u_{xxx}+\lambda (f)_x = 0
$$
Here $f = \frac{1}{4}tr(S^{2}_{x}), \quad \lambda =1. $

\end{document}